\documentclass{aa}

\usepackage{graphicx}

\def\note #1]{{\bf #1]}}

\def\dd{{\rm d}}
\def\cf{{\it cf.}}

\newlength{\figwidth}
\begin{document}

\title{  
On Variation of the Latitudinal Structure of the Solar Convection Zone
	}


\author{H. M.~Antia \inst{1}, S. M.~Chitre \inst{2} \and M. J.~Thompson \inst{3}}

\offprints{H. M.~Antia, \email{antia@tifr.res.in}}

\institute{Tata Institute of Fundamental Research, Homi Bhabha Road, 
   Mumbai 400005, India \and
Department of Physics, University of Mumbai, Mumbai 400098, India\and
   Space and Atmospheric Physics Group, The Blackett Laboratory, 
   Imperial College, London SW7 2BW, UK}
             
\date{Received ??, 2002; accepted ??}

\abstract{The latitudinal sound-speed structure of the Sun's convection 
zone gives insight into the physical processes occurring there, specifically 
the cellular convection and possibly the presence of magnetic fields. 
Using helioseismic data from the GONG network and MDI instrument on SOHO, 
we map the latitudinal acoustic structure of the convection zone from 
1995 to 2002. The temporally averaged structure confirms previous findings
of an excess in sound speed at the $10^{-4}$ level at 60 degrees latitude. 
There also appear to be some variations with time, with the peak in 
sound-speed asphericity at 60 degrees growing towards the maximum of solar
activity according to the MDI data. However, we present some 
evidence that such variation
may be associated with instrumental variation between the epochs before
and after SOHO was temporarily lost in 1998. Nonetheless, some genuine
temporal variation may be present, and we discuss the possible physical
causes of that.
}
\authorrunning{Antia, Chitre \& Thompson}
\titlerunning{Latitudinal structure of the solar convection zone}
\maketitle

\section{Introduction}

Analysis of the global p-mode oscillations of the Sun
reveals the Sun's internal structure and constrains the physics 
of the solar interior. In this paper we focus on the latitudinal 
variation of the structure of the solar convection zone and the
possible temporal variation of that structure during the rising phase
of solar cycle 23. The convection zone occupies the 
outer thirty per cent by radius of the Sun, as was deduced from 
helioseismic inversions for the Sun's radial hydrostatic structure 
(Christensen-Dalsgaard, Gough \& Thompson 1991; cf., Gough 1977; 
Ulrich \& Rhodes 1977). Such an investigation uses the mean multiplet
p-mode frequencies to deduce the spherically symmetric component of the 
Sun's structure. But with high-quality data such as from the 
Global Oscillation Network Group (GONG) and the Michelson Doppler
Imager (MDI) instrument on board
the SOHO satellite (Harvey et al.~1996; Scherrer et al.~1995) it has
become possible to make deductions also about the latitudinal variation of the 
structure from the dependence on the azimuthal order of the frequencies within 
multiplets (Gough et al.~1996; Antia et al.~2001).
What is measured is acoustic asphericity, i.e., the deviation from spherical 
symmetry of the propagation speed of the p-mode waves as a function of position.
The relevant wave speed is the adiabatic sound speed but possibly modified by magnetic effects
in regions of magnetic fields.
The acoustic 
structure of the convection zone may be expected to reflect 
the asphericity caused by the Sun's rotation, but also perhaps
the cellular 
convection and the presence of magnetic fields. At the resolution scale
at which the global helioseismic inversions can be undertaken, we are not
here concerned with granular or even supergranular scales. But the thermal
effects of giant cells or other such large-scale circulatory structures may be
detectable in the inferred latitudinal variation of sound speed, as perhaps
may also be a sufficiently strong large-scale magnetic field. In the present
paper we present results of inversions of a more extensive data set 
than was considered by Antia et al.~(2001). We also show that there is
some evidence for the temporal variation of the latitudinal
structure during the rising phase of the solar activity cycle. However,
one should be cautious because of possible instrumental variation during the
period of observations, and so we wish to examine such a possibility 
more closely.

\begin{figure*}[t]
\hbox to \hsize{\resizebox{\figwidth}{!}{\includegraphics{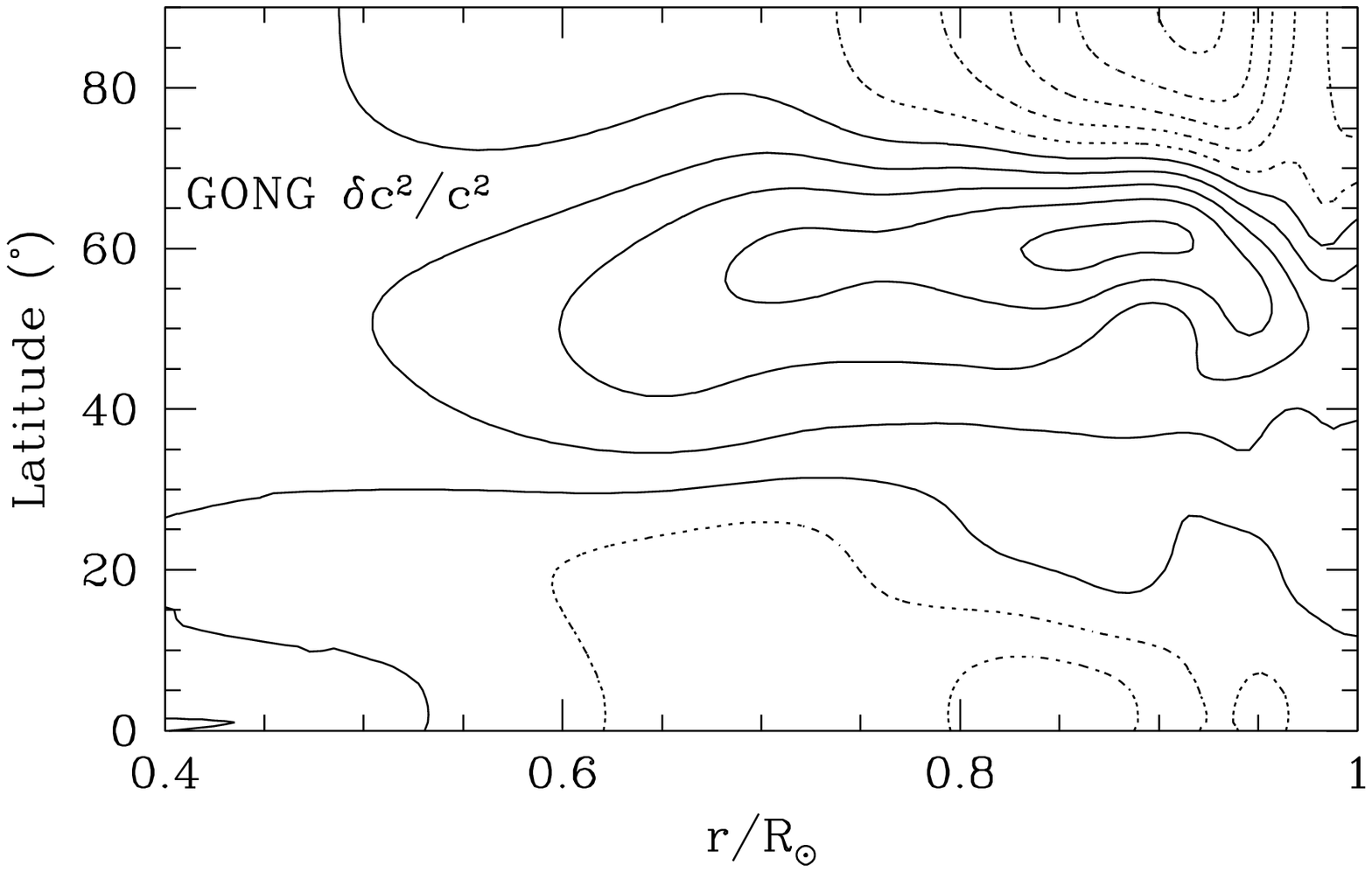}}
\quad\hfill \resizebox{\figwidth}{!}{\includegraphics{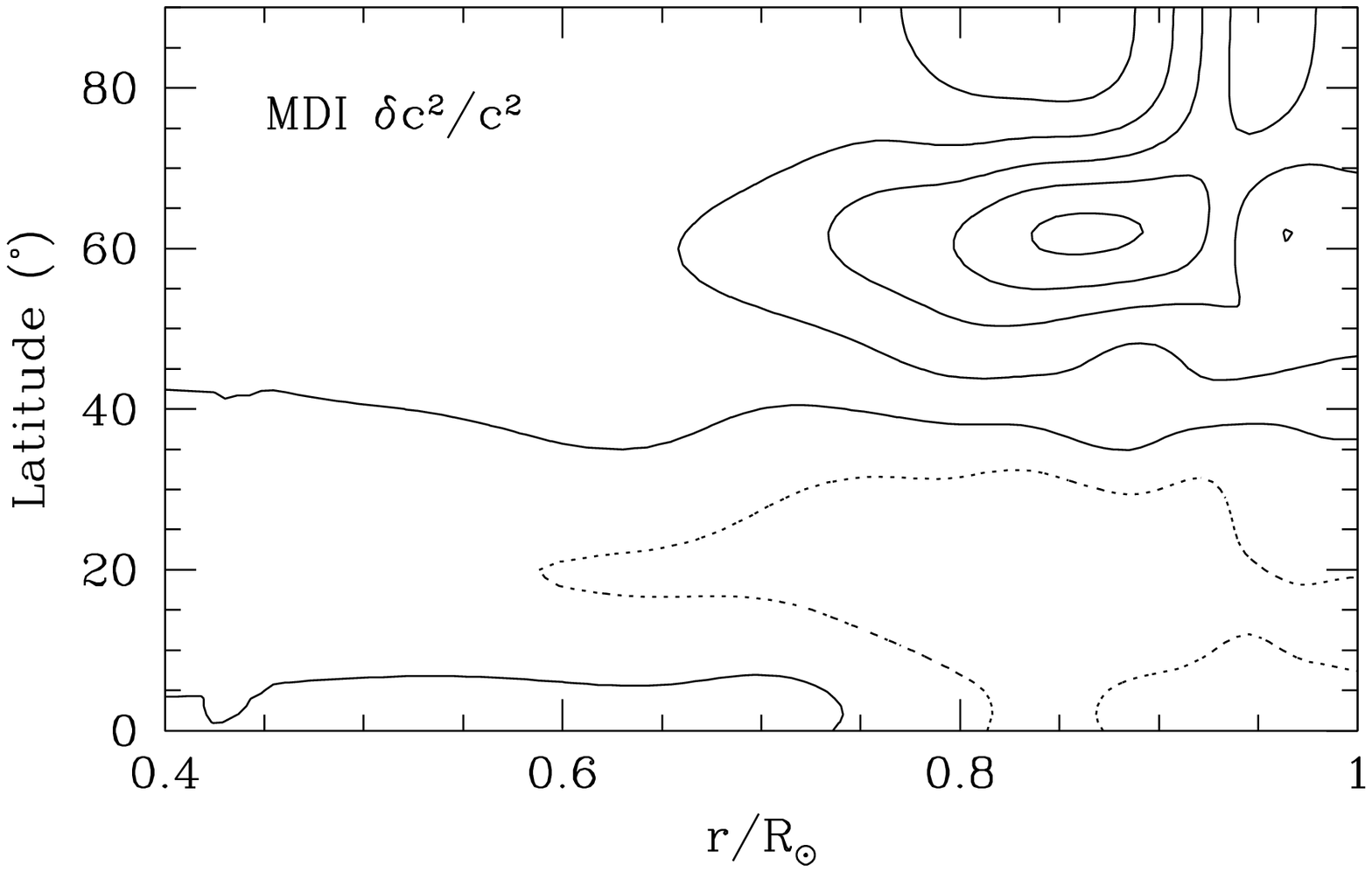}}}
\caption{Contour diagram of the temporal average of aspherical component of
squared sound speed 
as a function of radial distance and latitude using the GONG (left-panel)
and MDI (right-panel) data. The MDI data used are restricted to $\ell < 110$
(see Section 4).
Solid and dotted contours denote positive and negative values respectively.
Contours are drawn at interval of $2\times10^{-5}$.}
\label{fcs}
\end{figure*}

\section{Seismic data and inversions}

The helioseismic datasets comprise measures of frequencies (so-called
even a-coefficients) from GONG and MDI. 
The modes are identified by radial order $n$, spherical harmonic degree
$l$ and azimuthal order $m$. The frequencies $\nu_{n l m}$ 
are grouped in multiplets, 
corresponding to modes with all azimuthal orders (i.e.,
$m  = 0, \pm 1, \pm 2, \ldots, \pm l$) for given $n$ and $l$. In a 
spherically symmetric star, all frequencies in the multiplet would be 
equal, but rotation and asphericity raise this degeneracy.
Rather than report individual frequencies for every mode in the multiplet, 
the observers' data analysis procedure provides  
mean frequencies $\nu_{nl}$
and splitting coefficients $a_j(n,l)$ defined by
$$
\nu_{nlm} = \nu_{nl} + \sum_{j=1}^{j_{\rm max}}
a_j(n,l){\cal P}_j^{(l)}(m)\;,
$$
where the ${\cal P}_j^{(l)}(m)$ are orthogonal
polynomials in $m$ which are approximately equal to $l$ times Legendre
polynomial $P_j(m/l)$ when $l$ is large (e.g., Schou et al.~1994).

Odd splitting coefficients
$a_1$, $a_3$, etc. are used to infer the Sun's 
internal rotation (e.g., Schou et al.~1998): we shall not consider those
further here. The even coefficients contain information about the
Sun's aspherical structure, and here we use these coefficients, applying to
them a 2D regularised least-squares inversion technique as 
described by Antia et al.~(2001). In brief, the data are fitted with
an expression of the form
\begin{eqnarray}
\lefteqn{{l a_{2k}(n,l)\over \nu_{nl}}\ =}\nonumber\\
& &\left(2k+\frac{1}{2}\right)Q_{lk} \displaystyle\int_0^R\dd r
\int_0^\pi \sin\theta\,\dd\theta 
\left(K_{c^2,\rho}^{nl}(r){\delta c^2\over c^2}(r,\theta)\right.\nonumber\\
& &\displaystyle + \left. K_{\rho,c^2}^{nl}(r){\delta\rho\over\rho}(r,\theta)
\right)P_{2k}(\cos\theta)
 + Q_{lk}{F_k(\nu_{nl})\over E_{nl}}\;.\qquad
\end{eqnarray}
Here the known quantities are the kernels $K_{c^2,\rho}^{nl}$ and $K_{\rho,c^2}^{nl}$,
the factors $Q_{lk}$, and the mode inertias $E_{nl}$; and the unknowns are
the aspherical perturbations  
$\delta c^2$ and $\delta\rho$ to the squared sound speed and density 
respectively, and the surface terms $F_k$. The errors on
aspherical perturbations, $\delta c^2/c^2$
and $\delta\rho/\rho$ are estimated through a Monte-Carlo simulation.
The inversion procedure is repeated for different realisations of data,
with Gaussian noise added according to the
estimated errors on splitting coefficients,
to estimate the standard deviation in the values 
of $\delta c^2/c^2$ and $\delta\rho/\rho$ at each point.

The solar oscillations frequencies 
from the GONG project (Hill et al.~1996) were obtained from 108-day time series, while the
MDI data sets were derived from 72-day time series (Schou 1999).
We use data
sets that consist of mean frequencies and splitting coefficients.
In this work we are only concerned with even splitting coefficients
which are determined by the second order effect of rotation, asphericity
in solar structure or of a magnetic field.
We use 66 data sets from GONG
each covering a period of 108 days, starting from
1995, May 7 and ending on 2002, January 17
with a spacing of 36 days between consecutive data sets.
Thus each set overlaps with the two preceding and two following sets.
The MDI data sets (Schou 1999) consist of 28 non-overlapping
sets each obtained from observations taken over a   period of
72 days. The first set begins on 1996 May 1 and the last
set ends on  2002 March 30 with a gap in between during July 1998 to
January 1999 when contact with SOHO was temporarily lost.

Since the dominant north-south symmetric component of solar rotation rate is reliably
determined from odd splitting coefficients, we use these values to
calculate the second order effect of rotation as described by Antia et al.~(2000).
This second order contribution to even splitting coefficients is
subtracted from the observed coefficients to get the residuals which
could be due to the effect of
magnetic field or any other asphericity in solar structure.
In all results described in the following sections the rotational contribution
has been subtracted out. All these results have been obtained
through inversion of even coefficients $a_2,a_4,\ldots,a_{16}$.
The mean frequencies $\nu_{nl}$ are determined by
the spherically symmetric structure. Since this component has been
extensively studied we do not consider it in the present work.
The even splitting coefficients are sensitive only to the north-south symmetric
component of the structure variations and hence that is the only component that can
be determined with these data.

\section{Sun's latitudinal structure}

Several investigations of the medium-degree even splitting coefficients
in the last several years have pointed to a latitudinally dependent
spatial variation in sound propagation speed for the
solar p modes at a depth of about 40 Mm
beneath the surface (Antia et al.~2000; Dziembowski et al.~2000). The
magnitude of the effect is such that it could correspond to a relative
thermal perturbation at that depth of order $10^{-4}$, or a
magnetic field with rms field strength of roughly $50\,$kG.
Antia et al.~(2000)
speculate that such a field may be concentrated and retained at that
location by downward convective plumes (\cf, Li \& Sofia 2001).

The results of 2-D inversion by Antia et al.~(2001)
strikingly show that the
aspherical perturbation to the sound speed peaks at latitude of
about $60^\circ$, with a drop towards the pole.
These results have been updated using additional data that have become
available since then: Fig.~1 shows
the contours for the temporal average of the aspherical sound-speed perturbation
from the updated MDI and GONG data.
The MDI and GONG results are in reasonable agreement about the
location of the perturbation and its global behaviour. However, note that the period over which
the average is taken differs in the two panels because the GONG data
start a little earlier (in mid-1995) and there are also two periods
without data from MDI. 

\begin{figure}[t]
\resizebox{\figwidth}{!}{\includegraphics{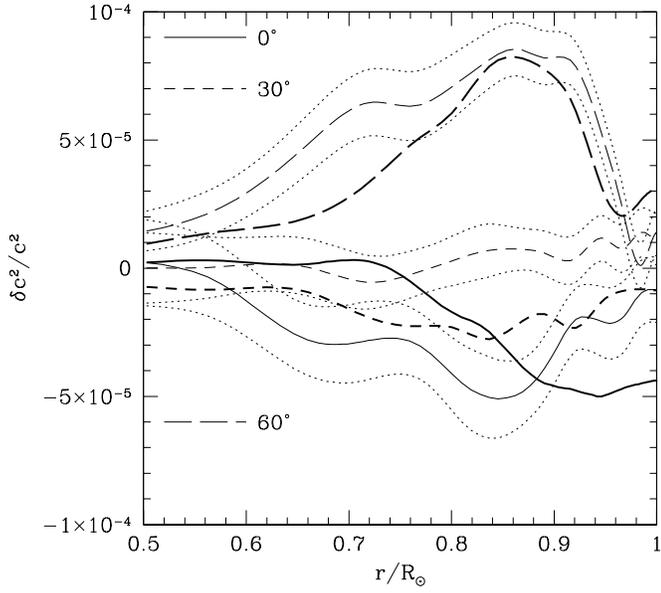}}
\caption{The aspherical component of squared sound speed as a
function of radius at a few selected latitudes as marked in the figure.
The thin lines are results using the GONG data with dotted lines showing
the error estimates, while the heavy lines are the MDI results.
For clarity the errors in MDI data are not shown, but they are comparable
to those in GONG data.}
\label{frad}
\end{figure}

\begin{figure}[t]
\resizebox{\figwidth}{!}{\includegraphics{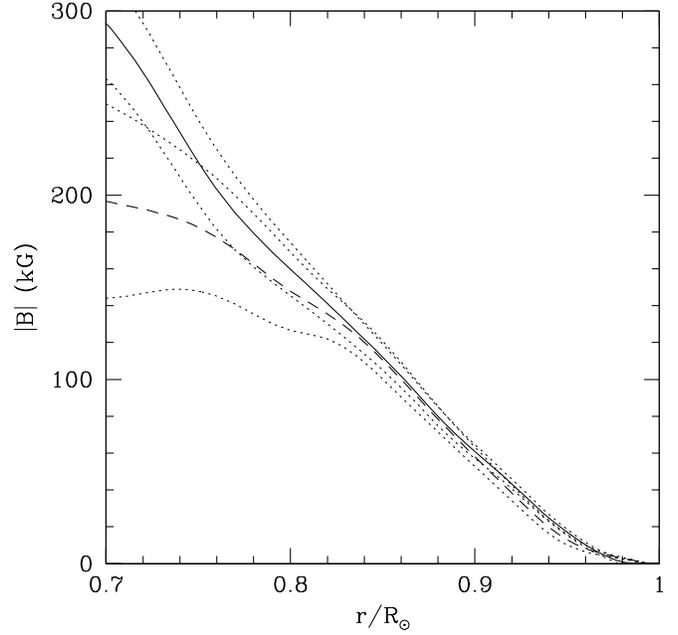}}
\caption{The estimated magnetic field at $60^\circ$ latitude assuming
that the inferred asphericity is due to magnetic field. The continuous
line shows the results from GONG data, while dashed line shows the
results using MDI data. In both cases the error limits are shown by
dotted lines.}
\label{fb60}
\end{figure}

\begin{figure}[t]
\resizebox{\figwidth}{!}{\includegraphics{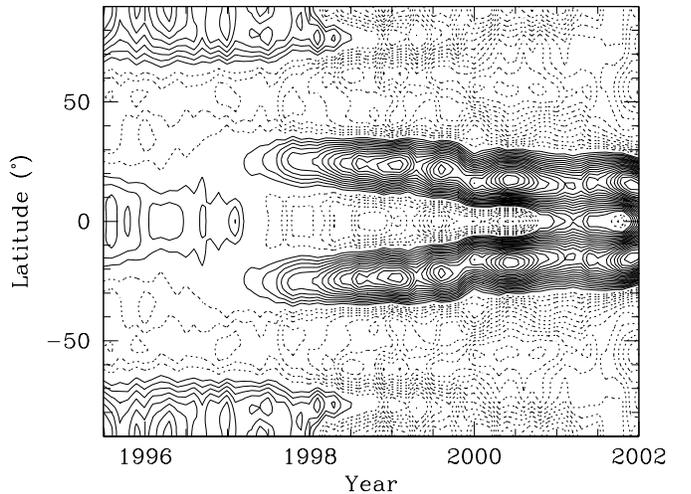}}
\caption{Contour diagram of the surface term in asphericity inversions
as a function of time and latitude using the GONG data.
The solid lines show positive contours and dotted lines show
negative contours.}
\label{fsurf}
\end{figure}

\begin{figure*}[t]
\resizebox{2.0\figwidth}{!}{\includegraphics{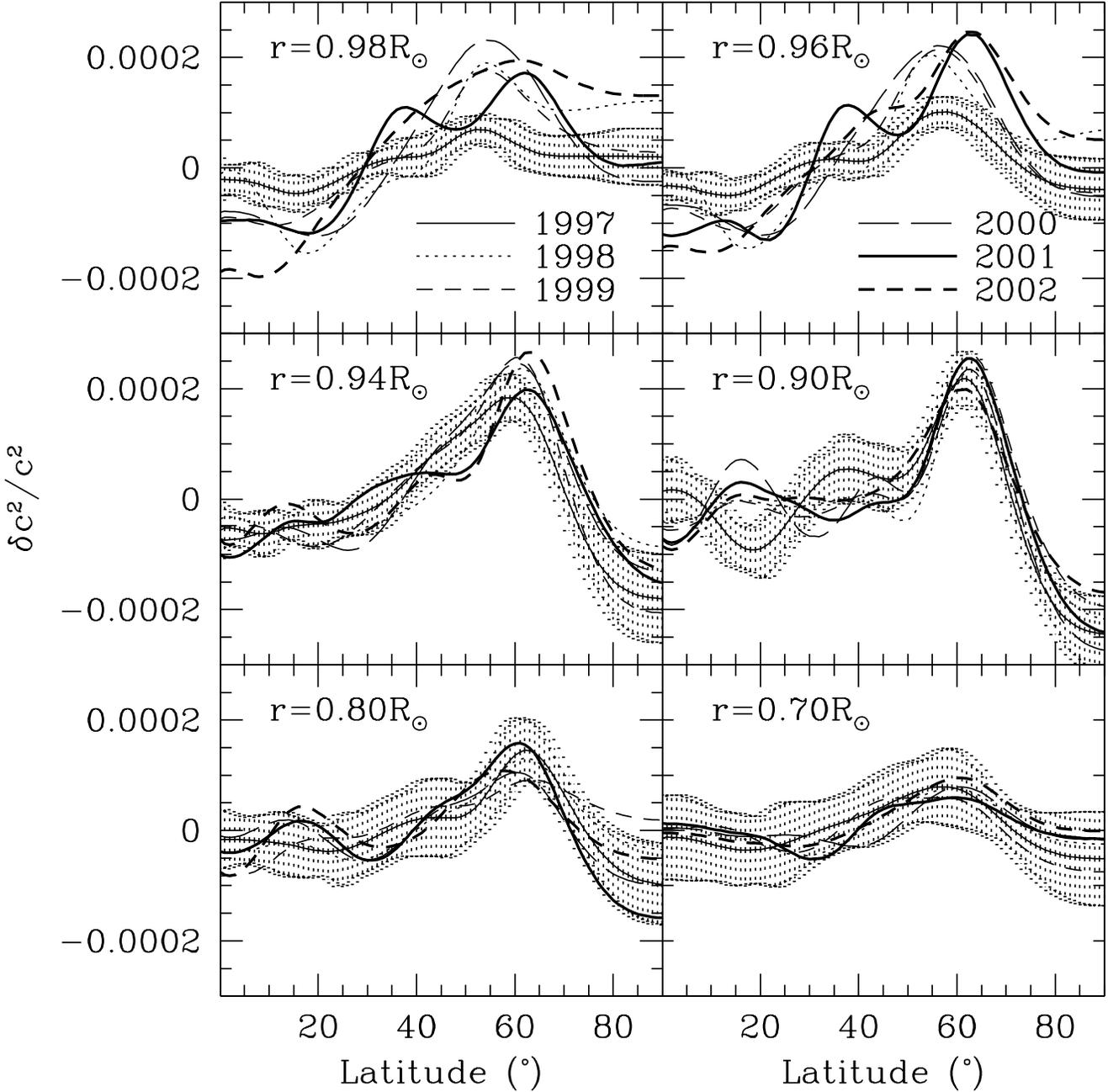}}
\caption{latitudinal cuts at fixed radius through the asphericity
$\delta c^2/c^2$ inferred from MDI data, for different radii as
indicated in each panel. The $1\sigma$ error bars are shown for one epoch
only as they are similar for all epochs.}
\label{fcuts}
\end{figure*}

Figure \ref{frad} shows the cuts through Fig.~1 at constant latitudes
from both GONG and MDI data. It is clear that the two results are
roughly consistent with each other. The cut through $60^\circ$ latitude
shows a clear peak around $r=0.9R_\odot$. We cannot tell from the seismic data
whether 
this is due to a thermal asphericity in the solar interior or to a magnetic field.
If it is due to a randomised magnetic field in the convection
zone, then this value may be taken as indicative of the order of magnitude of
the ratio $v_A^2/c^2$, where
$v_A$ is the Alfv\'en speed. From this we can calculate the
magnetic field: the results are shown in Fig.~\ref{fb60}.
Because of the increasing gas pressure the estimated magnetic field
keeps increasing with depth. Around the base of the convection zone
the magnetic field strength could be 200--300 kG, which is comparable
to the upper limit obtained by Antia et al.~(2000). The upper limit
of Antia et al.\ is applicable only to a field concentrated near
the base of the convection zone, while the magnetic field shown in
Fig.~\ref{fb60} is spread throughout the convection zone and hence
the earlier upper limit is not applicable.

The inversion results indicate that the
positive perturbation,
though peaked around $r=0.9R_\odot$,
extends through much of the convection zone. Although
the finite resolution of the inversion certainly spreads the features
around, it appears from a resolution study (Antia et al.~2001) that
the spread in depth seen here is more extensive than one would expect from a
perturbation wholly confined around $r=0.95R_\odot$ say.
The results appear to show that this peak extends below the convection
zone also, but this extension could be due to limited resolution of
the inversion process.
It may be pointed out that the peak of this perturbation has shifted
downwards as compared to what Antia et al.~(2000) found. This shift has 
arisen because,
unlike in our earlier work, in this work we have removed the surface
contribution ($Q_{lk}F_k(\nu_{nl})/E_{nl}$) during inversion. That shifts the peak a 
little downwards.  If this peak is due to magnetic field, the field strength
should be around 70 kG.

These results come from an inversion which attempts to isolate the
different contributions to Eq.~(1). There is a known excellent
correlation between the even splitting coefficients $a_{2k}$ and the surface
unsigned magnetic flux projected onto $P_{2k}(\cos\theta)$.
The latitudinal distribution of
surface flux varies over the solar cycle. This is reflected in the 
surface term that we fit in Eq.~(1). The surface term is shown 
in Fig.~\ref{fsurf},
which displays contours of constant surface term as a function of time
and latitude at a frequency of 3 mHz.
Note that scale and hence the contour spacing of
the surface term is arbitrary.
This figure shows distinct bands of positive surface
term moving towards the equator with time, which should coincide
with the bands in the magnetic butterfly
diagram. At high latitudes there is a hint of bands migrating
polewards, particularly, after 1998. This also has been seen in
magnetic features at the
solar surface (Leroy and Noens 1983; Makarov \& Sivaraman 1989b).

\section{Apparent temporal variations}

Clearly there is a substantial temporal variation in the even $a$ coefficients that
is associated with the near-surface variations of magnetic flux. This is taken up by
the surface term in Eq.~(1); see Fig.~\ref{fsurf}. Once this is removed, however, it is of
interest to see whether there is any residual temporal 
variation coming from the deeper layers.

We first study 
the development of acoustic asphericity by examining the cuts at different radii from the
MDI asphericity inversion results, at six different epochs (the beginnings
of 1997, 1998, 1999, 2000, 2001, 2002). 
Each of these results is obtained from MDI data set starting in the
beginning of the corresponding year and represents an average over
72 day period.
Fig.~\ref{fcuts} shows latitudinal cuts at fixed radii ($r/R = 0.98, 0.96, 0.94, 0.90, 0.80, 0.70$).
For the two highest locations, the asphericity clearly appears to
evolve, with a pronounced variance at high latitudes. In the
$60^\circ$ peak, there is a substantial difference between 1997 and
later years, and the peak strengthens and broadens approaching 2000.
The slightly deeper cuts show the peak sharpening but the polar values are
very steady. Much deeper in the convection zone (bottom two panels) the
amplitude is much reduced, but it is still somewhat larger than one
might expect if the underlying perturbation were strictly confined in the
layers around e.g. $0.95R_\odot$. Thus its spread to greater depths may
be to some extent real. It is difficult to make out any systematic
temporal variation in asphericity from this figure as even in the upper
layers the variation is only of the order of 2--3$\sigma$.

\begin{figure}[t]
\resizebox{\figwidth}{!}{\includegraphics{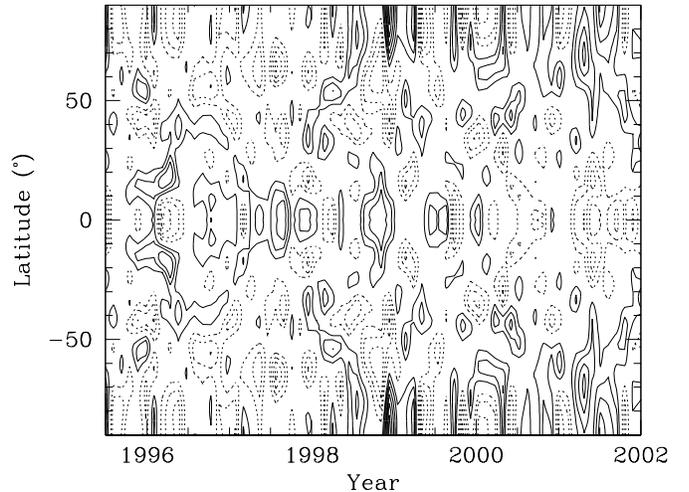}}
\caption{Contour diagram of the residual in aspherical component of the
squared sound speed after subtracting the temporal average,
as a function of time and latitude at $r=0.96R_\odot$ using the GONG
data. The solid lines show positive contours and dotted lines show
negative contours. Contours are drawn at interval of $2\times10^{-5}$.}
\label{fgong}
\end{figure}

Because there is an apparent temporal variation we should look at its origin more closely.
Fig.~\ref{fgong} shows as a contour plot the results for aspherical
component of squared sound speed
obtained using GONG data at $r=0.96R_\odot$
as a function of latitude and time.
To see the temporal variations
more clearly, the figure shows the residuals
obtained after subtracting the temporal average at each latitude.
There is no evident pattern in the residuals.
Similar results are obtained at other depths.  

\begin{figure*}[t]
\hbox to \hsize{\resizebox{\figwidth}{!}{\includegraphics{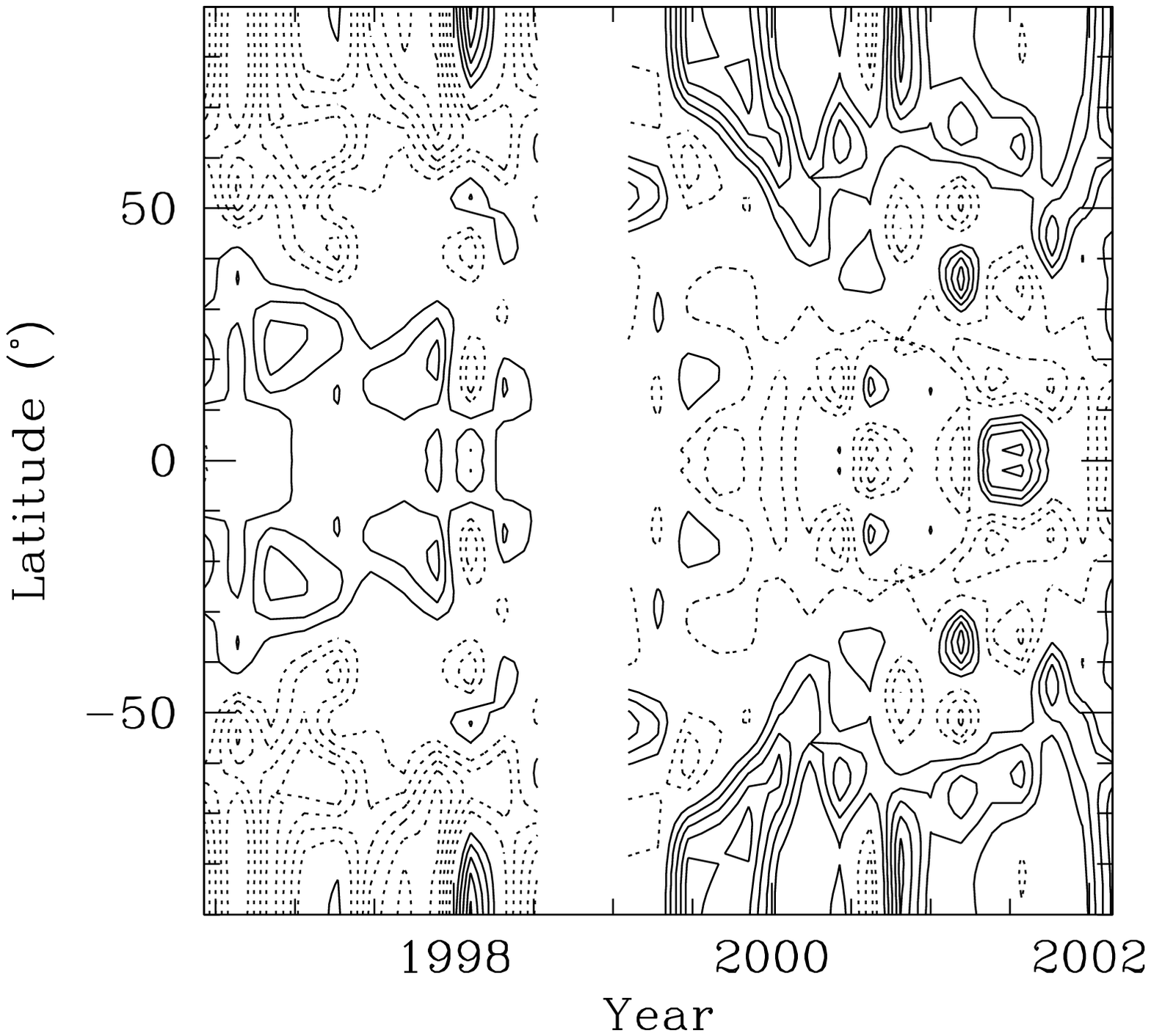}}
\quad\hfill \resizebox{\figwidth}{!}{\includegraphics{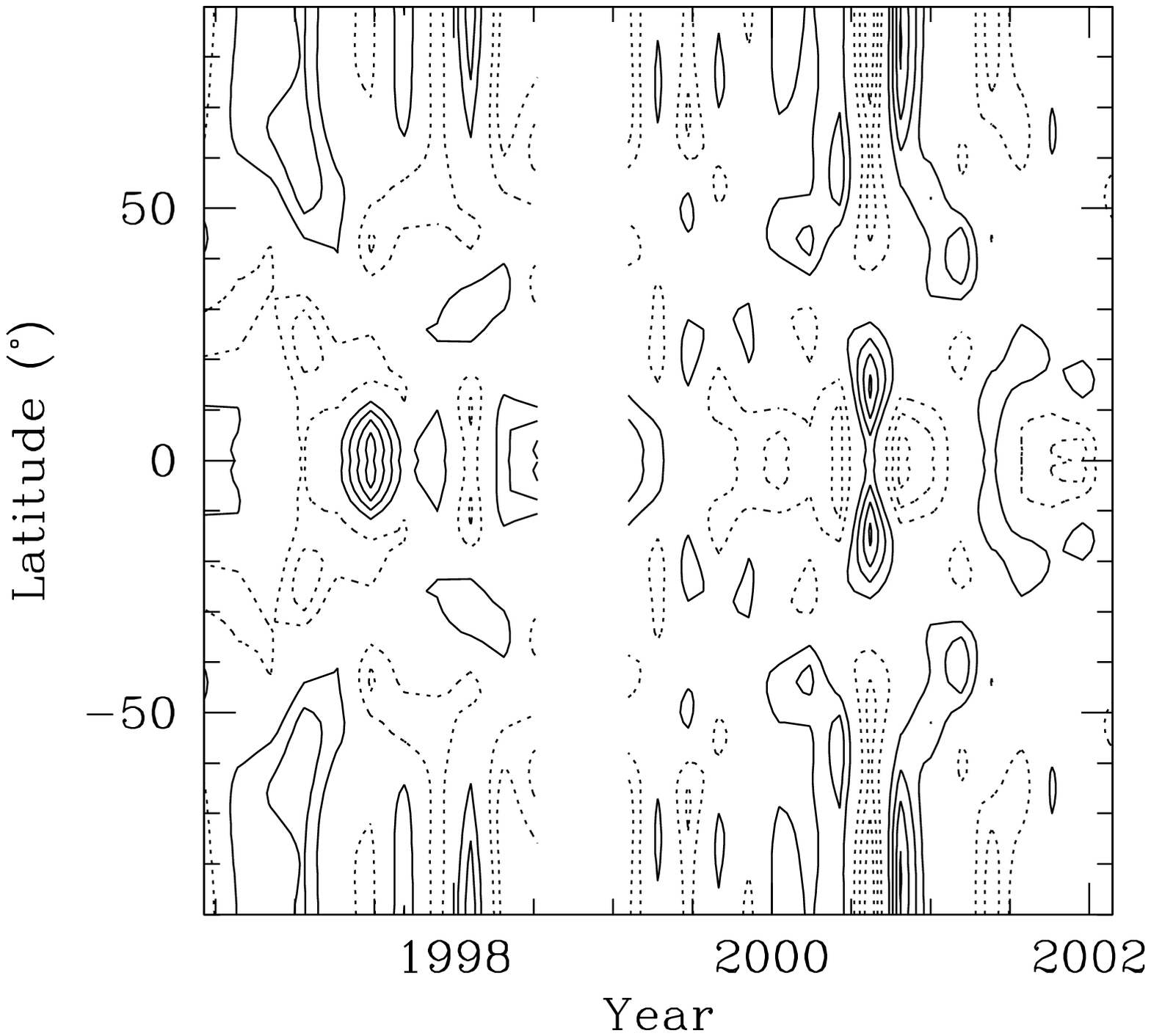}}}
\caption{Contour diagram of the residual in aspherical component of
squared sound speed after subtracting the temporal average,
as a function of radial distance and latitude using the MDI data.
The solid lines show positive contours and dotted lines show
negative contours. Contours are drawn at interval of $2\times10^{-5}$.
The left panel shows the results using all modes with $1.5\le\nu\le3.5$
mHz, while the right
panel shows the results using modes with $1.5\le\nu\le3.0$ and
$\ell<110$.}
\label{fmdi}
\end{figure*}

Fig.~\ref{fmdi} similarly shows the residuals in sound speed at $r=0.96R_\odot$, but using
the MDI data. The left panel which displays the results
using full data shows a distinct temporal variation. This is consistent 
with what was seen in Fig.~\ref{fcuts}, but is in contrast
to the GONG results (Fig.~\ref{fgong}). Antia (2002) has pointed out that 
most of the temporal variation occurs between July 1998 and
January 1999, when the contact with SOHO satellite was lost. This
period is
marked by the gap in the contour diagram. Before the gap
there is positive asphericity at low latitudes and negative values
at high latitudes. This is reversed after the gap and in fact there
is very little systematic variations in the results before or after
the gap. 
It is possible the Sun has had some interesting transition while 
the MDI instrument was not operating. This hypothesis gets little
support from the GONG data, however, which does not appear to 
show any unusual trend during this period.
We must therefore consider the possibility that the apparent variation is
due to systematic errors introduced by instrumental variations during the
recovery of the SOHO satellite. We suspect the high-degree data may be
responsible: the right panel of Fig.~\ref{fmdi}, which shows results using only modes
with $\ell<110$, displays no particular temporal variation in
asphericity. In fact, the temporal average for the MDI data shown
in Fig.~\ref{fcs} is also obtained using only modes with $\ell<110$.
If all modes are used then the peak is much more pronounced with maximum
value of more than $2\times10^{-4}$
which is double the peak height
in Fig.~1. As a result, in the following results
we have used only the restricted mode set from MDI.

\begin{figure*}[t]
\hbox to \hsize{\resizebox{\figwidth}{!}{\includegraphics{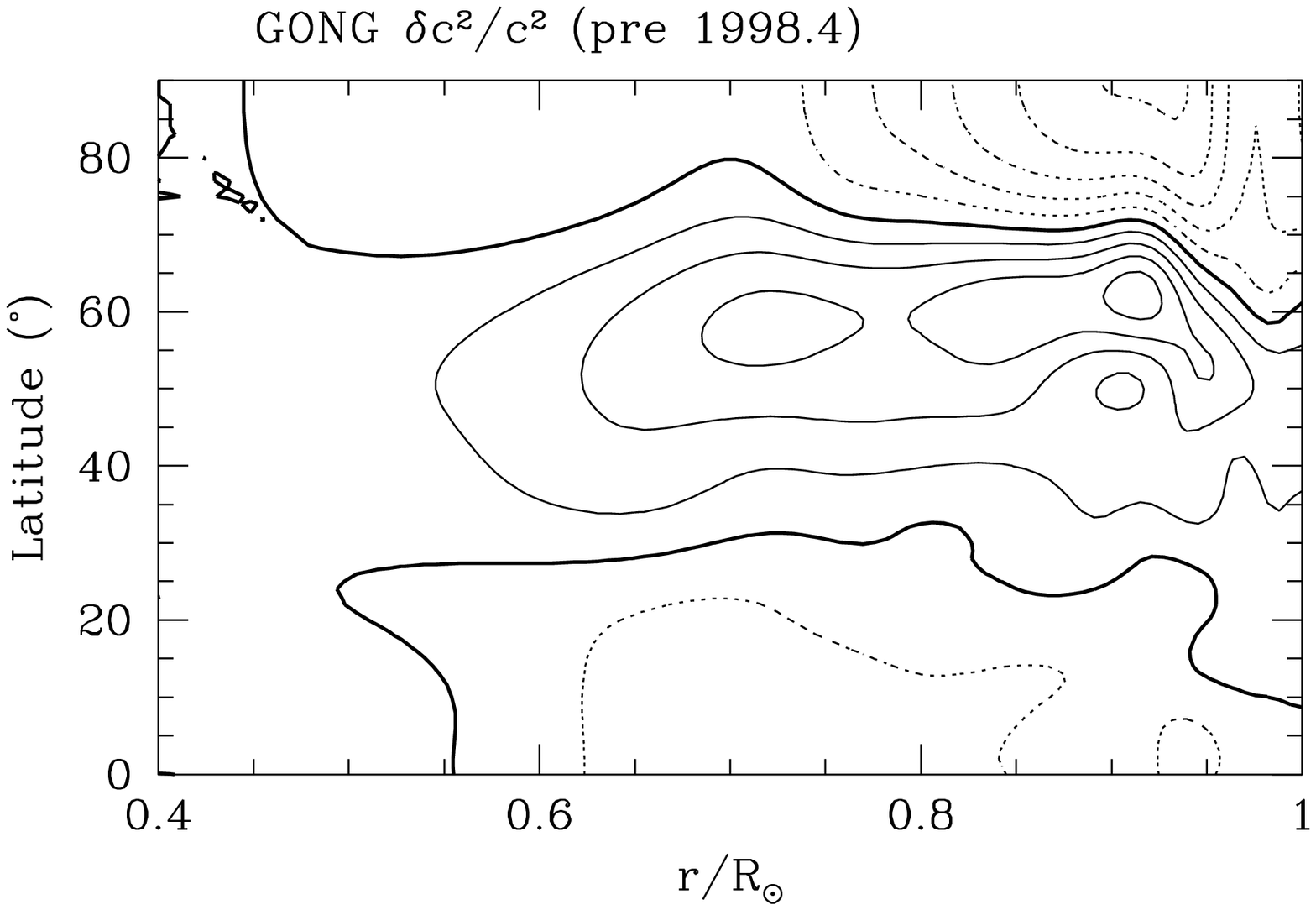}}
\quad\hfill \resizebox{\figwidth}{!}{\includegraphics{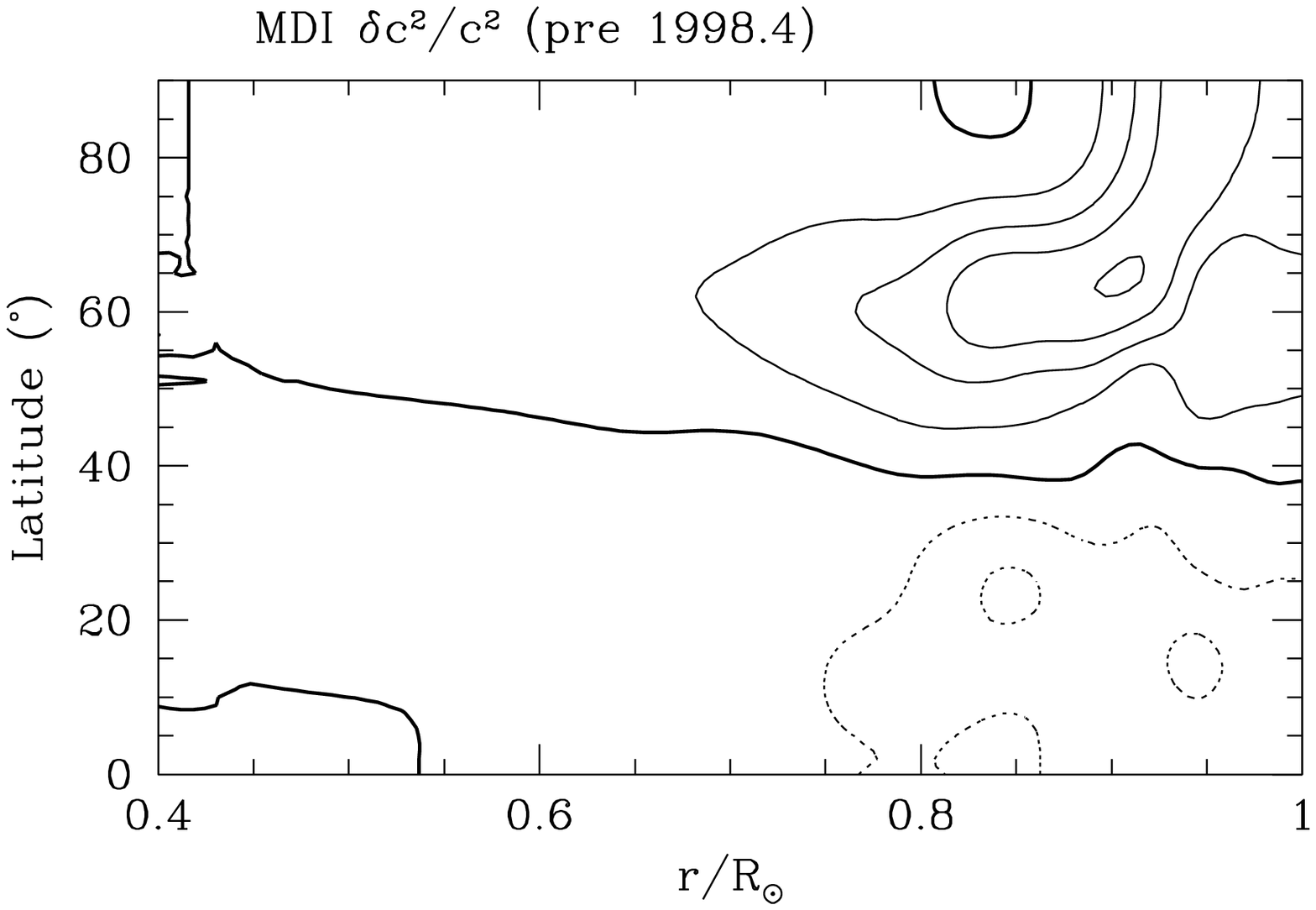}}}
\bigskip
\hbox to \hsize{\resizebox{\figwidth}{!}{\includegraphics{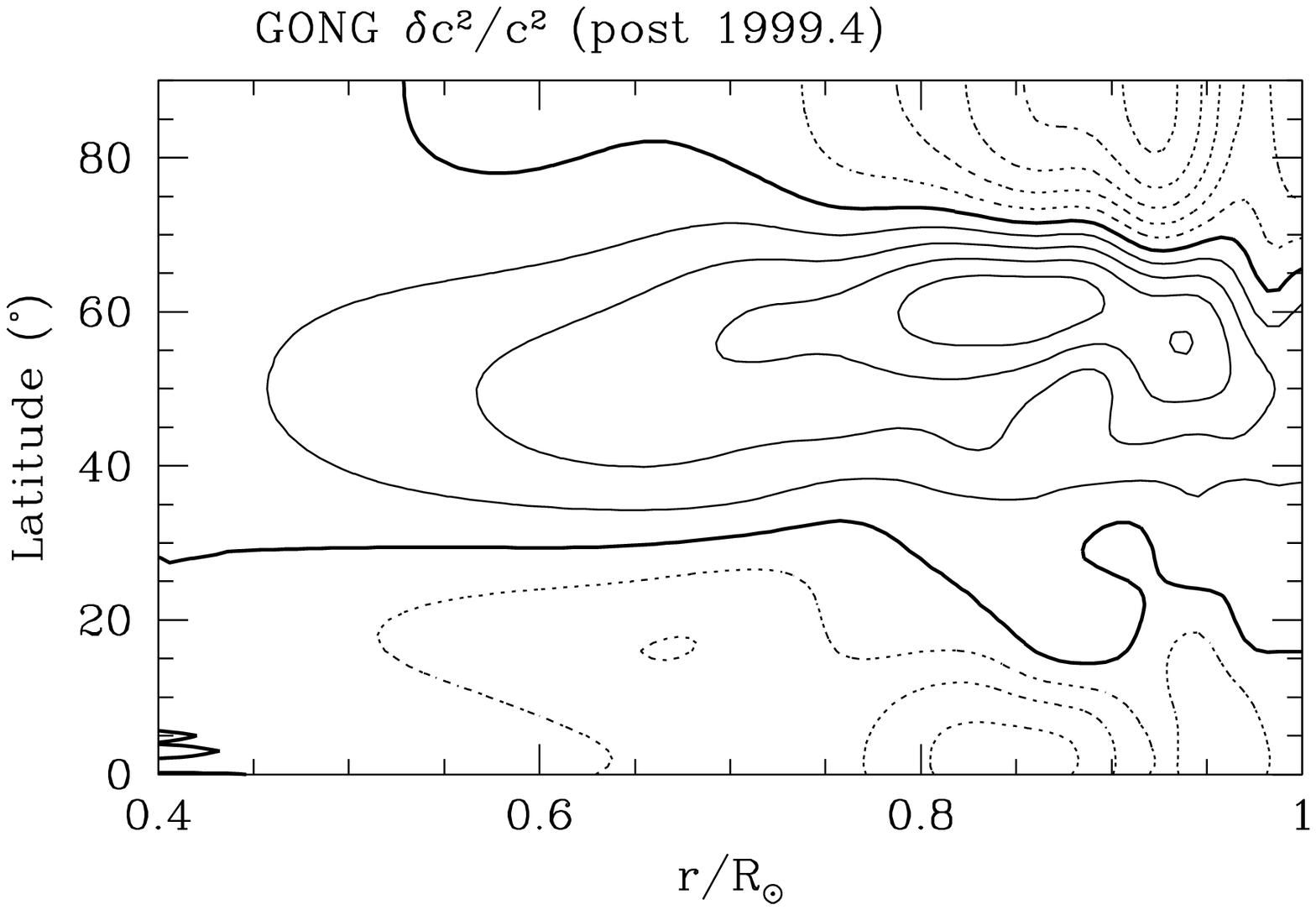}}
\quad\hfill \resizebox{\figwidth}{!}{\includegraphics{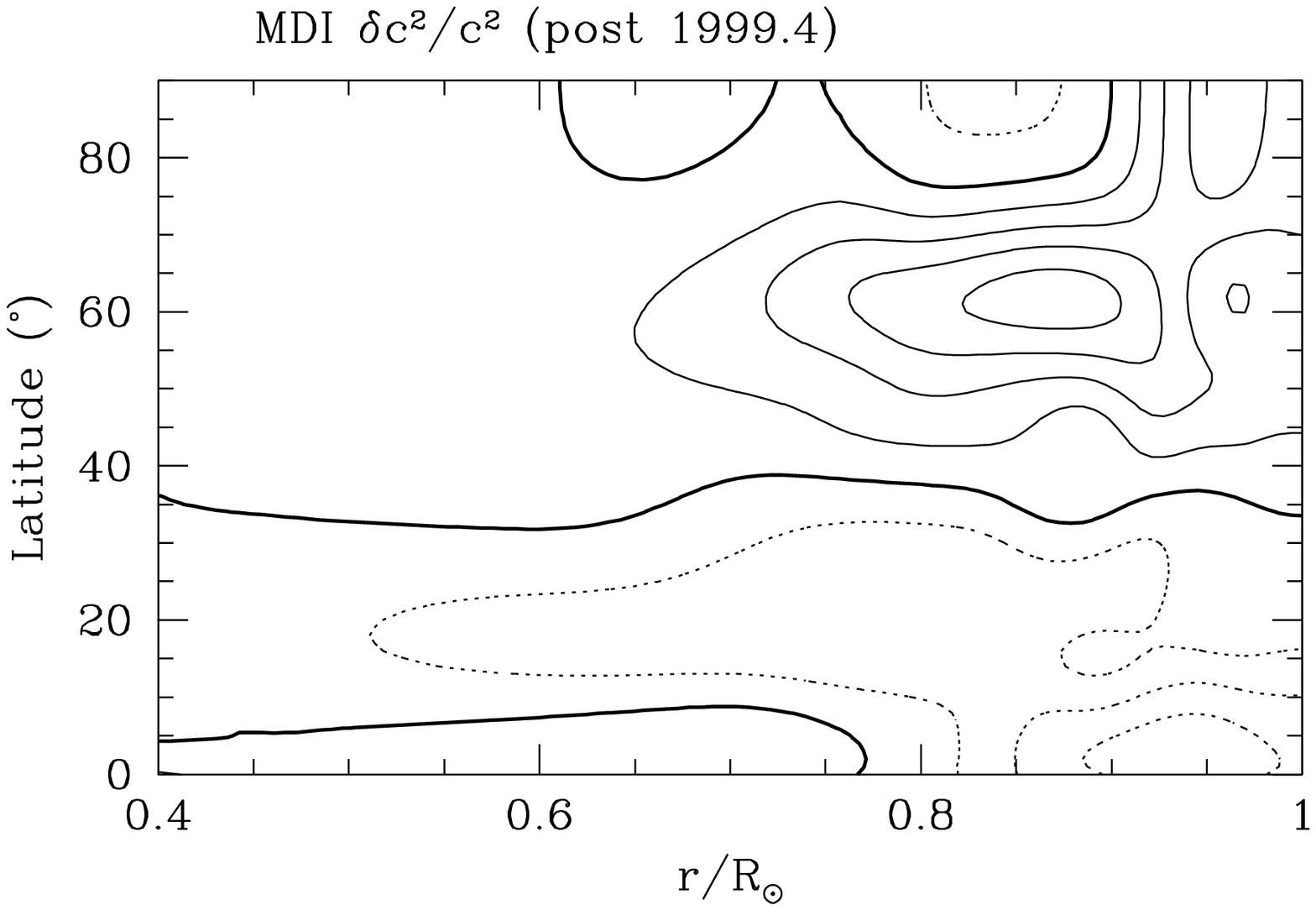}}}
\bigskip
\hbox to \hsize{\resizebox{\figwidth}{!}{\includegraphics{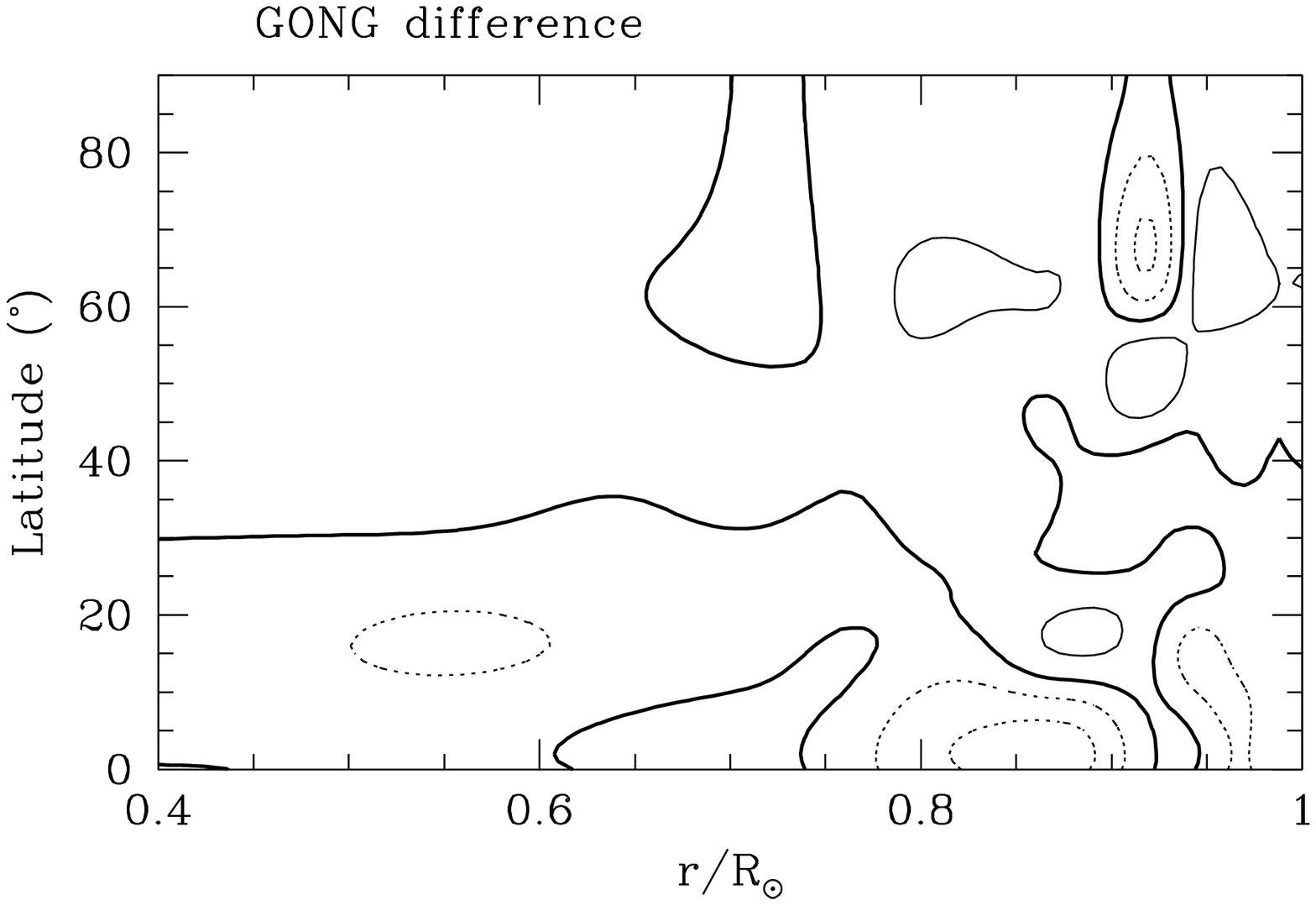}}
\quad\hfill \resizebox{\figwidth}{!}{\includegraphics{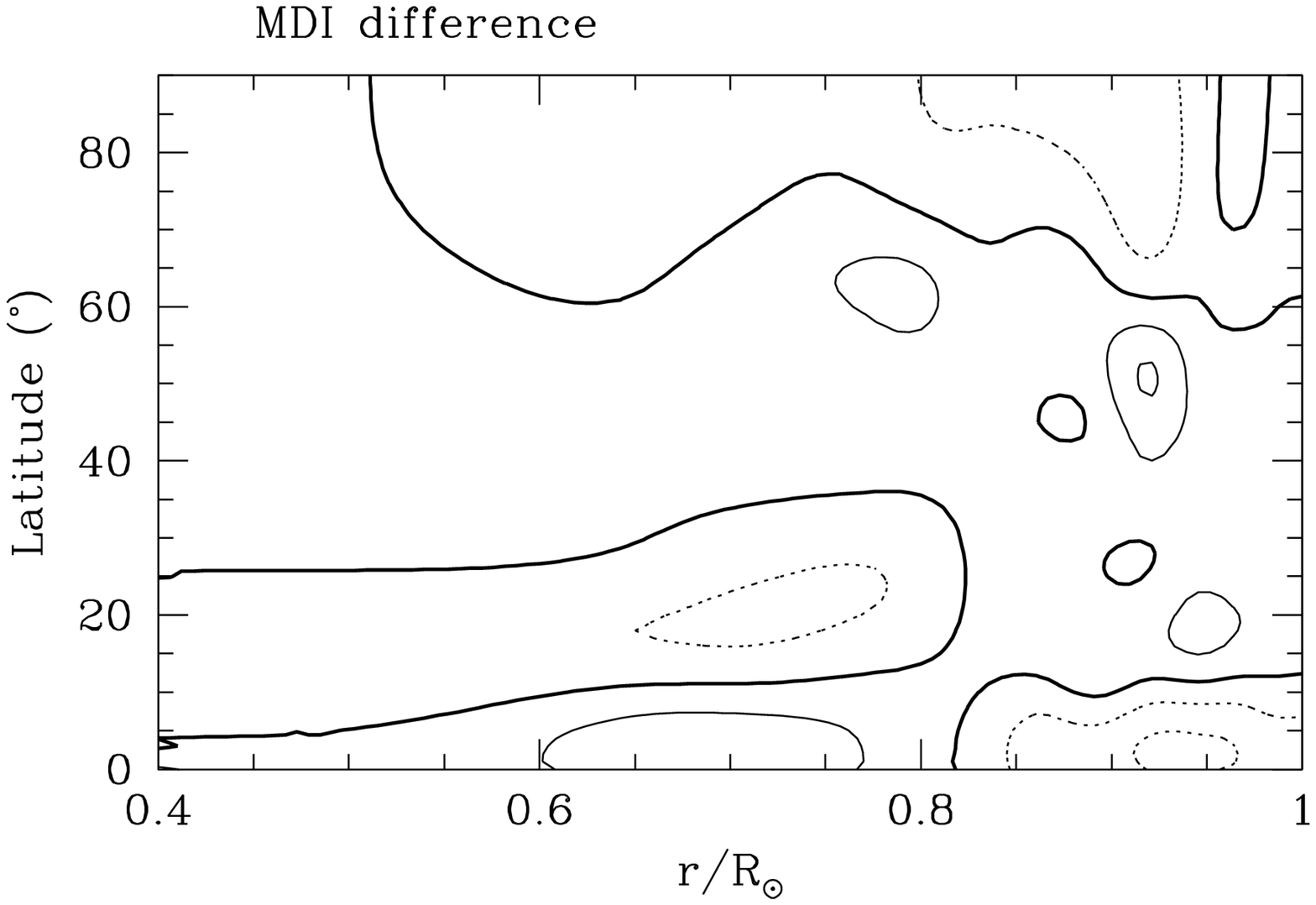}}}
\caption{Contour diagram of the temporal average of aspherical component of
squared sound speed
as a function of radial distance and latitude.
The solid lines show positive contours and dotted lines show
negative contours. Contours are drawn at interval of $2\times10^{-5}$.
The thick contour in these figures mark the zero level.
The temporal average using GONG data before 1998.4 (top-left),
after 1999.4 (middle-left), and MDI data before 1998.4 (top-right),
after 1999.4 (middle-right) are shown. The two panels
at the bottom show differences in acoustic asphericity
between the high and low activity
averages for GONG (bottom-left) and MDI (bottom-right).}
\label{ftime}
\end{figure*}

With a view to studying the possibility of small temporal variation in inferred
asphericity in sound speed, we divide the data sets into two parts:
a low activity period before 1998.4 and a high activity period after 1999.4,
and take temporal averages separately over these parts.
The results are shown in Fig.~\ref{ftime}. There is a slight increase in
asphericity in high activity set as compared to that in low activity set
in both GONG and MDI results. Fig.~\ref{ftime} also shows the difference
between the high and low activity periods. The
increase appears at a level of $\sim 2\times 10^{-5}$ in some regions and is
spread over a wide region, though the maximum change appears to
occur around $r=0.9R_\odot$ at high latitude as well as in the equatorial
region.
By comparison, the typical errors
in low latitude region inside the convection zone is about $10^{-5}$
which is half the contour spacing. The errors increase with depth
and latitude. The significance of this increase in asphericity is
not very clear, but with the availability of data over a complete solar
cycle it may be
possible to confirm whether there is indeed an increase in asphericity
with activity.

\section{Discussion}

The helioseismic inversions for asphericity in sound speed show a
distinct peak at a latitude of $60^\circ$ inside the convection zone.
These inferences raise a number of questions. What is so special
about $60^\circ$ latitude? Perhaps it is the boundary between two
oppositely directed meridionally circulating cells, or at least the
poleward limit of a low-latitude cell. If the perturbation to the
sound speed is indeed caused by magnetic field, as we have suggested
(Antia et al.~2000), then the flux could accumulate at this latitude in
a convergent lane between the two cells.
We can envisage two possibilities. Earlier studies of surface magnetism
emphasised the poleward migration of prominences and faculae from
around this location (e.g., Waldmeier 1981;
Makarov \& Sivaraman 1989a,b) and led to a
picture where this latitude had surface flows diverging from it. In that
case, the cells could act to dredge dynamo-generated flux from the lower
convection zone into this latitude. On the other hand, ring diagram analyses
of local helioseismic data (Basu et al.~1999; Haber et al.~2000) find poleward meridional
circulations at low latitudes, in the outer 20 Mm or so that they
can image. Small-scale fields could be generated by the alpha effect 
operating in the upper convection zone, or could come from the decay of 
old sunspots and active regions. The meridional
flows could sweep up this near-surface flux 
and collect it like jetsam in a downflow lane at $60^\circ$ latitude.

A further point of interest about $60^\circ$ latitude is that this is where
the near-surface shear changes sign, according to MDI helioseismic measurements:
the angular velocity increases with depth beneath the surface at low
latitudes whereas it decreases with depth at high latitudes (Schou et al.
1998; Corbard \& Thompson 2002). This could reflect the role
of differently signed meridional circulation flows 
in the transport of angular momentum
on either side of $60^\circ$.
Furthermore, 
the time-dependence of rotation at high latitudes is perhaps another
manifestation of the interplay between magnetic field and flows.
The variation in rotation rate in the upper convection zone shows
a pattern with bands of faster and slower than average rotation
rate moving towards the equator with time at low latitudes
(Howe et al.~2000; Antia \& Basu 2000) while at high latitudes these
bands appear to be moving towards the pole (Antia \& Basu 2001).
The transition between equatorward and poleward movement occurs
around a latitude of $50^\circ$.
The near-surface rotation at high latitudes, as revealed by
helioseismology, has been slowing down until 1999 during the rising phase
of the present cycle (Schou 1999; Antia \& Basu 2001). At high latitudes,
the rotation rate decreases with time, reaches
a minimum and then starts increasing again. The epoch at which
the minimum occurs is a function of latitude: it gradually
shifts to later times as one goes to higher latitudes.
One might envisage that
it is the open magnetic field lines that
slow the rotation. As the open flux retreats glacier-like
back to high latitudes, it releases its brake on the rotation at a given
latitude and so at that latitude the rotation rate springs back up.

One of the outstanding problems in solar physics is to identify the
location of a plausible mechanism responsible for simultaneous variations
of the Sun's luminosity, radius, effective temperature and p-mode
frequencies with the activity cycle. 
The seat of the solar dynamo is generally believed to be located in the 
tachocline region beneath the convection zone where the mechanism for solar 
luminosity variation could also reside. However, it has been argued by 
Balmforth et al.~(1996) that an agent like a magnetic field concentrated at
the base of the solar convection zone is unlikely to be directly responsible for the observed cyclic modulations; nor can the purely thermal changes occurring
in the near-surface layers of the Sun account for the observed irradiance 
variations. Any deeply seated driving mechanism would yield values of
the quantity $W = (dR/R)/(dL/L)$, which measures the relative changes in 
solar radius and luminosity, somewhere between $0.2$ and $0.5$. The current
estimates of $W$ seem to suggest much lower values, of order $0.02$
(Emilio et al.~2000). In fact, these results from satellite observations 
indicate that the principal cause of the solar luminosity variation is 
unlikely to be located very deep inside the Sun.

Helioseismic inversions have revealed the existence of a shear layer 
below the photosphere,
extending to depths of about 50 Mm 
below the photosphere, with its
peak occurring at 30 Mm depth, in the midst of the helium
ionisation zone. (The above values depend somewhat on latitude, 
the depth of shear layer reducing with increasing latitude.)
Dikpati et al.~(2002) have explored the role of this subphotospheric
radial shear layer in driving a solar dynamo, and find that a differentially rotating
shear layer is capable of generating toroidal magnetic fields of up to
1 kG strength. Antia et al.~(2000) had invoked the pumping of downward
moving plumes (downdrafts) to concentrate toroidal magnetic field in
the outer convection zone and to enhance its strength to values approaching 
70 kG at depths around 70 Mm below the surface.
This takes place
at depths around 70 Mm below the surface, where some sort of 
equipartition is achieved between the magnetic energy density 
$\langle B^2 / 8\pi\rangle$ and the kinetic energy density of downdrafts
$\langle \rho w^2\rangle$.
Note, the velocities of the order of 1 km/s adopted here, refer to
the downdraft motions which have velocities several times those
given by the standard mixing length theory for convective
elements (cf., Mestel 1999).
This process, of course, does not exclude the 
generation of a large fraction of the toroidal magnetic field in the 
solar interior, with the help of differential rotation in the tachocline.
We are merely pointing out that part of the toroidal field in the convection
zone may be generated in the sub-surface shear layer itself, where the 
poloidal field could be reinforced by some kind of alpha effect. This would
then be immediately available for further generation of the toroidal field
by the action of strong subsurface radial shear. 

Earlier it was noted by Antia et al.~(2000) and Dziembowski et al.~(2000) that
residual splitting coefficient $a_2$ showed a striking peak around 
$r=0.96R_\odot$. If this peak should result solely from the presence of 
large-scale magnetic field at depths of order 30 Mm, the inferred 
magnetic field strength comes out to be 20--30 kG. But with the removal
of surface term, this peak shifts downwards to 70 Mm depth and
required magnetic field strength would be about 70 kG.

It is readily seen from Fig.~1 that the contours of constant
$\delta c^2/c^2$ in both the GONG and MDI data show a
pronounced radial extension around the $60^\circ$ latitude peak
situated at a depth of about 70 Mm.
Even though a comparison of the contours of constant $\delta c^2/c^2$
in the pre-1998.4 and post 1999.4 GONG and MDI data sets displayed
in Fig.~8 barely shows a detectable temporal variation, the difference
in acoustic asphericity between the high and low activity averages
shown in the bottom panels of Fig.~8 has some indication of a
possible temporal variation of asphericity around a depth of
$0.1R_\odot$ at both equatorial and higher latitudes.
It may therefore be tempting to speculate that there is a reservoir of magnetic 
energy stored in the outer $0.1R_\odot$ of the convection zone which
perhaps undergoes some temporal variation. Such a time-varying 
storage of magnetic energy might even play a 
role in driving the solar activity cycle. 
We suggest that the solar irradiance variation is probably due to the 
periodic release of such a stored magnetic energy in the outer layers of the 
Sun, resulting from an interplay between the solar magnetic field, subsurface
radial shear and pumping by convective downdrafts.

\begin{acknowledgements}
This work  utilises data obtained by the Global Oscillation
Network Group (GONG) project, managed by the National Solar Observatory,
which is
operated by AURA, Inc. under a cooperative agreement with the
National Science Foundation. The data were acquired by instruments
operated by the Big Bear Solar Observatory, High Altitude Observatory,
Learmonth Solar Observatory, Udaipur Solar Observatory, Instituto de
Astrofisico de Canarias, and Cerro Tololo Inter-American Observatory.
This work also utilises data from the Solar Oscillations
Investigation/ Michelson Doppler Imager (SOI/MDI) on the Solar
and Heliospheric Observatory (SOHO).  SOHO is a project of
international cooperation between ESA and NASA.
SMC is grateful to DAE-BRNS for support under the Senior
Scientist Scheme
and to Ian Roxburgh for supporting his visit to Queen Mary, University
of London under the Leverhulme Trust Visiting Professorship scheme.

\end{acknowledgements}


\begin{thebibliography}{}

\bibitem[]{a02} Antia, H. M. 2002,
in proc. IAU Coll.\ 188: Magnetic Coupling of
the Solar Atmosphere, ESA SP-505 (astro-ph/0208339)

\bibitem[Antia \& Basu(2000)]{ab00}
Antia, H. M., \& Basu, S. 2000, ApJ, 541, 442

\bibitem[Antia \& Basu(2001)]{ab01}
Antia, H. M., \& Basu, S. 2001, ApJ, 559, L67

\bibitem[]{ant00} Antia, H. M., Chitre, S. M. \& Thompson, M. J. 2000, 
      A\&A, 360, 335


\bibitem[]{ant01} Antia, H. M., Basu, S., Hill, F., Howe, R., Komm, R. W. \& 
      Schou, J. 2001, MNRAS, 327, 1029

\bibitem[]{bal96}
Balmforth, N. J., Gough D. O., Merryfield W. J. 1996, MNRAS, 278, 437

\bibitem[]{bat99}
Basu, S., Antia, H. M., \& Tripathy, S. C. 1999, ApJ, 512, 458

\bibitem[]{jcd91} Christensen-Dalsgaard, J., Gough, D. O. \& Thompson, M. J. 1991,
      ApJ, 378, 413 

\bibitem[]{cor02} Corbard, T. \& Thompson, M. J. 2002, Solar Phys., 205, 211

\bibitem[]{dik02} Dikpati,  M., Corbard, T., Thompson, M. J. \& Gilman, P. A. 2002,
      ApJ, 575, L41 

\bibitem[]{dz00} Dziembowski, W. A., Goode, P. R., Kosovichev, A. G. \& 
      Schou, J. 2000, ApJ, 537, 1026

\bibitem[Emilio et al.(2001)]{MDI2001}
      Emilio, M., Kuhn, J. R., Bush, R. I., \& Scherrer, P. 2000, ApJ, 543, 1007

  \bibitem[]{dog77} Gough, D. O. 1977,
      in {\it Proc. IAU Colloq. No. 36: The energy balance and 
      hydrodynamics of the solar chromosphere and corona},
      ed.\ R. M. Bonnet \& P. Delache
      (G. de Bussac, Clairmont-Ferrand) 3 

  \bibitem[]{dog81} Gough, D. O. 1981, 
      in Variations of the Solar Constant,
      ed. S. Sofia (NASA CP 2191) 185

\bibitem[]{dog96} Gough, D. O., et al.~1996,
      Sci.,  272, 1296

\bibitem[]{hab01} Haber, D. A., Hindmann, B. W., Toomre, J., Bogart, R. S. \&
     Hill, F. 2001,
     in Proc. SOHO 10 / GONG 2000 Workshop,
     ed A. Wilson (ESA SP-464) 213

\bibitem[]{har96} Harvey, J. W., et al.~1996,
      Sci.,  272, 1284

\bibitem[]{hi96}
Hill, F. et al.~1996, Sci., 272, 1292

\bibitem[Howe et al.(2000)]{how00}
Howe, R., Christensen-Dalsgaard, J., Hill, F., Komm, R. W.,
Larsen, R. M., Schou, J., Thompson, M. J. \& Toomre, J. 2000a, ApJ, 533
, L163

\bibitem[1983]{le83}
Leroy, J. -L., \& Noens, J. -C. 1983, A\&A, 120, L1

   \bibitem[]{li01} Li, L. H. \& Sofia, S. 2001, 
      ApJ, 549, 1204

   \bibitem[]{mak89a} Makarov, V. I. \& Sivaraman, K. R. 1989a, 
      Sol. Phys., 119, 35

   \bibitem[]{mak89b} Makarov, V. I. \& Sivaraman, K. R. 1989b, 
      Sol. Phys., 123, 367

\bibitem[]{mes99} Mestel, L. 1999, Stellar Magnetism, Clarendon press
(Oxford) p.~306

\bibitem[]{sch95} Scherrer, P. H., et al.~1995,
      Sol. Phys., 162, 129

   \bibitem[]{sch99} Schou, J. 1999, 
      ApJ, 523, L181

   \bibitem[]{sch94} Schou, J., Christensen-Dalsgaard, J. \& Thompson, M. J. 1994,
      ApJ, 433, 389

\bibitem[]{sch98} Schou, J., et al.~1998, ApJ, 505, 390


  \bibitem[]{ulr77} Ulrich, R. K. \& Rhodes, E. J. 1977,
      ApJ, 218, 521 

   \bibitem[]{wal81} Waldmeier, M. 1981, 
      Sol. Phys., 70, 251

\end{thebibliography}
\end{document}